\documentclass[letter]{aa}
\usepackage{natbib}
\bibpunct{(}{)}{;}{a}{}{,}
\usepackage{txfonts}

\newcommand{\msunyr}{\ensuremath{\mathit{M}_{\odot}{\rm yr}^{-1}}}   
\newcommand{\kms}{\ensuremath{{\rm km\,s^{-1}}}}                   
\newcommand{\msun}{\ensuremath{\mathit{M}_{\odot}}}   
\newcommand{\lsun}{\ensuremath{\mathit{L}_{\odot}}}                  
\newcommand{\rsun}{\ensuremath{\mathit{R}_{\odot}}}                  

\newcommand{\lstar}{\ensuremath{\mathit{L}_{\star}}}                 
\newcommand{\mdot}{\ensuremath{\dot{M}}}                             
\newcommand{\rstar}{\ensuremath{\mathit{R}_{\star}}}                 
\newcommand{\teff}{\ensuremath{\mathit{T}_{\rm eff}}}                
\newcommand{\vinf}{\ensuremath{\upsilon_{\infty}}}                          
\newcommand{\K}{\ensuremath{\mathrm{K}}}                 

\newcommand{\tauross}{\ensuremath{\tau_{\mathrm{ross}}}}                 

\newcommand{\vwind}{\ensuremath{\mathit{v}_\mathrm{wind}}}

\begin{document}

\title{Early-time spectra of supernovae and their precursor winds: the luminous blue variable/yellow hypergiant progenitor of SN 2013cu} 

\author{Jose H. Groh}
\institute{
Geneva Observatory, Geneva University, Chemin des Maillettes 51, CH-1290 Sauverny, Switzerland; \email{jose.groh@unige.ch}
}
\authorrunning{Groh et al.}
\titlerunning{The LBV/YHG progenitor of SN 2013cu}

\date{Received  / Accepted }

\abstract{We present the first quantitative spectroscopic modeling of an early-time supernova that interacts with its progenitor wind. Using the radiative transfer code CMFGEN, we investigate the recently-reported 15.5~h post-explosion spectrum of the type IIb SN 2013cu. For the first time, we are able to directly measure the chemical abundances of a SN progenitor and find a relatively H-rich wind, with H and He abundances (by mass) of $X=0.46\pm0.2$ and $Y=0.52\pm0.2$, respectively. The wind is enhanced in N and depleted in C relative to solar values (mass fractions of $8.2\times10^{-3}$ and $1.0\times10^{-5}$, respectively). We obtain that a dense wind/circumstellar medium, with a mass-loss rate of  $\mdot\simeq3\times10^{-3}~\msunyr$ and $\vwind\simeq100~\kms$, surrounds the precursor at the pre-SN stage. These values are lower than previous analytical estimates,  although we find  $\mdot/\vinf$ consistent with previous work. We also compute a CMFGEN model to constrain the progenitor spectral type and find that the high $\mdot$ and low $\vwind$ imply that the star had an effective temperature of $\simeq~8000~\K$ immediately before the SN explosion. Our models suggest that the progenitor was either an unstable luminous blue variable or a yellow hypergiant undergoing an eruptive phase, and rule out a WR star. We classify the post-explosion spectra at 15.5~h as XWN5(h) and advocate for the use of the prefix `X' (eXplosion) to avoid confusion between post-explosion, non-stellar spectra with those of massive stars. We show that the progenitor spectral type is significantly different than the early post-explosion spectral type owing to the huge differences in the ionization structure before and after the SN event. We find the following temporal evolution: LBV/YHG  $\rightarrow$ XWN5(h) $\rightarrow$ SN IIb. Future early-time spectroscopy in the UV will give access to additional spectroscopic diagnostics and further constrain the properties of SN precursors, such as their metallicities.}

\maketitle

\section{\label{intro}Introduction} 
\defcitealias{galyam14}{G14}

Core-collapse supernovae (SN) are the final act in the evolution of stars more massive than about 8--9~\msun. Determining the progenitors of these explosive events and how massive stars are linked to the different SN types are topics of major significance for several fields of Astrophysics.  

Numerous techniques have been employed to constrain the nature of SN progenitors. For instance, direct detection of progenitors in pre-explosion images has yielded the determination of the photospheric properties of the progenitors, such as effective temperature and luminosity \citep[e.g.,][]{smartt09a}. However, pre-explosion imaging provide weak constraints on the progenitor chemical abundance, mass loss, and wind speed. So far, mass loss immediately before the SN explosion 
has been investigated mostly in SN IIn using spectroscopy and photometry several days after their  discovery (e.g, \citealt{smith07,pastorello07,kiewe12,ofek14b,moriya14a}). 

Recent progress in observational techniques now allow for rapid-response spectroscopic observations of SNe within a day of detection \citep[][hereafter G14]{galyam14}. This allows the study of early phases when the SN shock front has not yet reached spatial scales of $10^{14}$~cm. Depending on the progenitor's wind density and SN shock front velocity, these early-time SN observations may probe epochs early enough that the dense parts of the progenitor wind and circumstellar medium (CSM) have not yet been overrun by the SN shock front. 

This was the case for the type IIb SN 2013cu (iPTF13ast) reported by \citetalias{galyam14}. The spectroscopic observations obtained 15.5~h after first detection reveal surprising features that apparently resemble those seen in Wolf-Rayet (WR) stars of the WN6(h) subtype. Based on this spectrum, \citetalias{galyam14} suggested a WR-like progenitor with evidence of  H in its wind, thus being consistent with the SN IIb classification obtained from later spectra. These authors also found evidence of enhanced mass loss prior to the SN explosion, with a mass-loss rate of $\mdot\sim 0.01~\msunyr$.

Here we present the first radiative transfer modeling of spectra of the early-time interactions of a SN with its dense progenitor wind. Our goal is to constrain the progenitor properties of SN 2013cu, such as its chemical composition, mass-loss rate, and wind velocity, from detailed modeling of spectroscopic features. As we discuss below, one of our key findings is that the progenitor spectral type does not correspond to the spectral type detected in rapid-response, post-explosion spectra. In the case of 2013cu, our models indicate that the progenitor was a luminous blue variable (LBV) or yellow hypergiant (YHG).

\section{\label{model}Early-time modeling of SN 2013cu}

We employ the non-local thermodynamical equilibrium, spherical,  line-blanketed, atmospheric/wind radiative transfer code CMFGEN \citep{hm98} to investigate the early post-explosion properties of SN 2013cu via spectroscopic modeling. At this point, we perform no hydrodynamical modeling. Our models are specified by the location of the inner boundary ($R_\mathrm{in}$), bolometric luminosity ($L_\mathrm{SN}$),  constant progenitor mass-loss rate (\mdot) and wind velocity ($v_\mathrm{wind}$), and  H, He, C, N, and O abundances. Solar abundances are used for P, S, and Fe. 

We assume a steep density gradient with a scale height of $0.007 R_\mathrm{in}$ to simulate the shock layer, which is joined smoothly to the non-shocked progenitor wind characterized by a density profile $\rho \propto r^{-2}$. We assume diffusion approximation at the inner boundary, which has a  density $\rho_{in}$ that is adjusted to obtain a Rosseland optical depth (\tauross) of 50 at $R_\mathrm{in}$. This ensures that the photons are thermalized at all wavelengths at the inner boundary (see also \citealt{dessart05a}, although their models are for non-interacting SN at later epochs). These conditions mimic those of  SN shock breakout that occurs in the progenitor wind/circumstellar medium (CSM; see \citealt{ofek10,chevalier11,moriya14b}), as proposed by \citetalias{galyam14} for SN 2013cu. We also assume that no energy is generated in the progenitor wind, that time-dependent effects are negligible, and that the medium in unclumped. By fitting an observed spectrum, we are able to constrain the following properties, using the diagnostics in parenthesis: \\
-- progenitor $\mdot$ (strength of the H$\alpha$ line);\\ 
-- progenitor $\vwind$ (width of the narrow component of emission lines);\\
-- $L_\mathrm{SN}$ and $R_\mathrm{in}$ (ionization structure and absolute flux assuming a distance $d$ to the SN);\\
--  progenitor chemical abundances (H$\alpha$, \ion{He}{ii} $\lambda$5411, \ion{C}{IV} $\lambda\lambda$5801--5810, \ion{N}{iv}  $\lambda\lambda$7109--7123).

We are aware that our post-explosion modeling has several simplifications and caveats that may affect the observables. First, we consider a stationary wind for the progenitor with constant mass-loss rate. Second, the hydrodynamics of the inner wind may be modified by the huge radiation field from the SN shock breakout as well as the radiation from the wind--SN interaction \citep{ofek10,nakar10,chevalier11,rabinak11,moriya11}, with the inner wind being accelerated \citep{fransson13}. Radiation hydrodynamical simulations of SN interacting with a dense CSM show that acceleration of the inner wind may occur depending on the wind density and SN luminosity  \citep{moriya11}. However, CMFGEN cannot handle non-monotonic outflows at the moment. Third, the radiation field and level populations may be affected by time dependence, since the physical conditions rapidly change within the first day after shock breakout \citep[e.g.,][]{dessart11a,dessart11b,gezari08}. However, as long as the photosphere (defined as the layer where \tauross=2/3) is located in the progenitor wind and the SN shock front is deep at high optical depths, our working model should allow us to obtain, for the first time, realistic quantitative information about SN progenitor winds via spectroscopic modeling.

\begin{figure}
\resizebox{0.995\hsize}{!}{\includegraphics{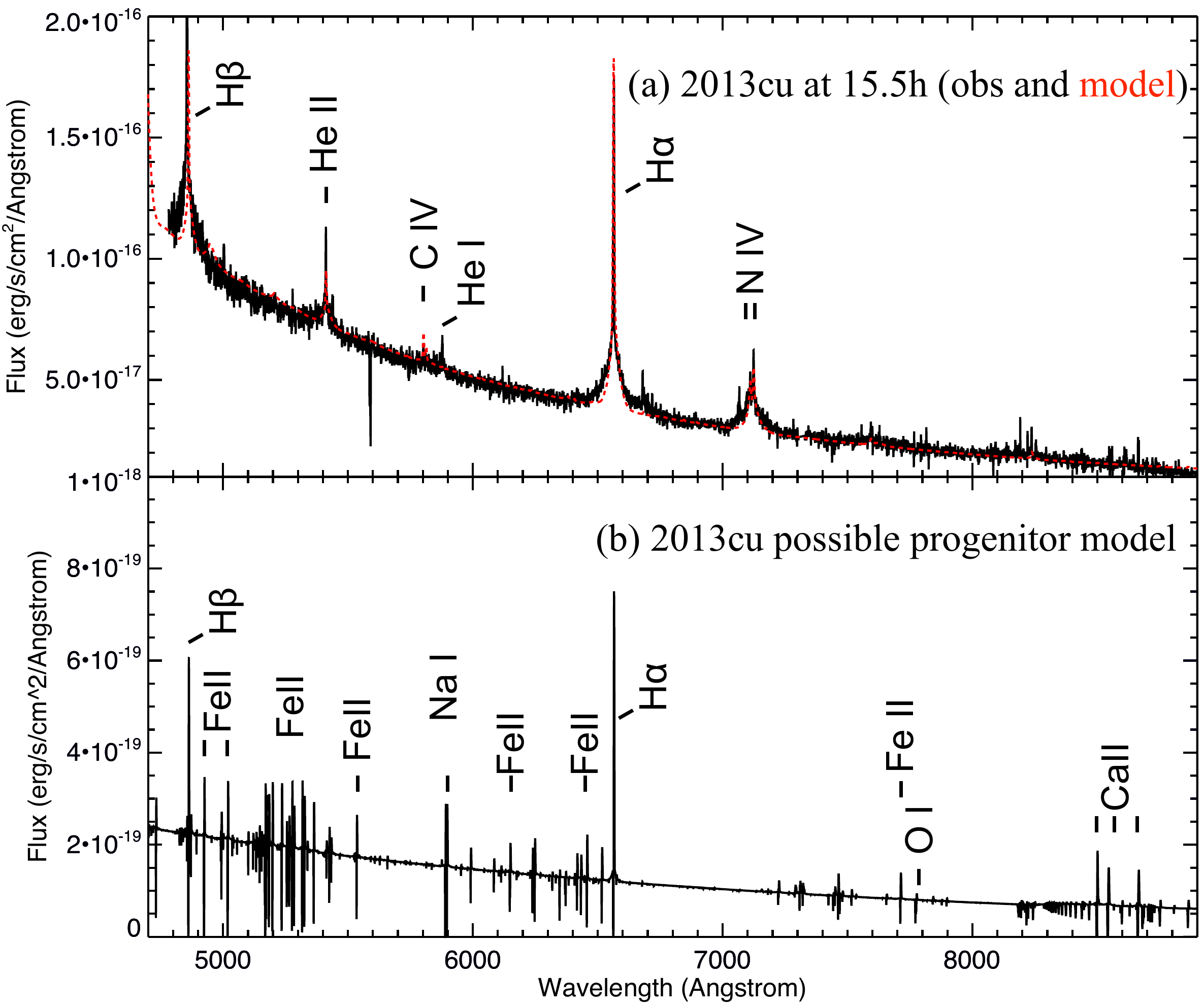}}\\
\caption{\label{fig1}{ (a) Model of the  spectrum of SN 2013cu at 15.5h after the explosion (red-dashed line) and the respective observations from G14 (black). The strongest features are labelled. (b) Example of a possible LBV pre-explosion spectrum of the progenitor of SN 2013cu. Notice the huge difference in pre- and post-explosions morphology due to very different ionization conditions before and after the SN.}}
\end{figure}

Figure \ref{fig1}a shows the observed optical spectrum of 2013cu at 15.5~h after the explosion (G14) and our best-fitting CMFGEN model. We obtain that a dense wind/CSM, with $\mdot\simeq3\times10^{-3}~\msunyr$ and  $\vwind\simeq100~\kms$, surrounds the progenitor at the time of explosion. These values are lower than the  analytical estimates of \citetalias{galyam14} ($1\times10^{-2}~\msunyr$ and  $500~\kms$,  respectively), although our determination of $\mdot/\vinf$ is roughly consistent with this previous study. Our best-fitting model has  $L_\mathrm{SN}=1\times10^{10}$~\lsun\ and $R_\mathrm{in}=1.5\times10^{14}$~cm, which is consistent with the dense portions of the wind not being overrun by the SN shock front.  

The early-time spectral morphology of SN 2013cu is well reproduced by our model (Fig. \ref{fig1}a). We can reproduce reasonably well the strength and shape of most features, with a narrow component and  broad emission wings that are caused by electron scattering (and not by the wind velocity field). The blue wings of the Balmer lines are underestimated by our models, which may be a sign that the inner wind is accelerated \citep{fransson13}. The strength of the \ion{He}{i} lines is also underestimated by our models, although a larger value of $R_\mathrm{in}$ would better reproduce the  \ion{He}{i} lines. This uncertainty in  \ion{He}{i} lines yields an error of a factor of 2 in $\mdot$ and $L_\mathrm{SN}$. It also hampers our ability to precisely constrain the relative H and He abundances. 

Yet, our modeling allows us to directly determine, for the first time, the chemical composition of a SN progenitor wind. We find a significant amount of H and He, with abundances (by mass) of $X\approx0.46 \pm 0.2$ and $Y\approx0.52\pm0.2$, respectively. Future model improvements could possibly reduce these uncertainties significantly. The wind is enhanced in N and depleted in C relative to solar values (mass fractions of $8.0\times10^{-3}$ and $1.0\times10^{-5}$, respectively). Assuming a solar CNO content yields a depleted O  mass fraction of  $1.6\times10^{-4}$.  These values are consistent with a progenitor that lost a sizable fraction of the H envelope and presents fully CNO-processed material at the surface, confirming the suggestions from \citetalias{galyam14}. It is important to note that the chemical composition of the SN ejecta will generally be significant different from the progenitor wind abundance. This is particularly true in the case of SN IIb progenitors, where the H-rich layer that will be later ejected in the SNe contains small amounts of mass ($\sim0.1~\msun$; \citealt{dessart11b,hachinger12}).

\begin{figure}
\resizebox{0.995\hsize}{!}{\includegraphics{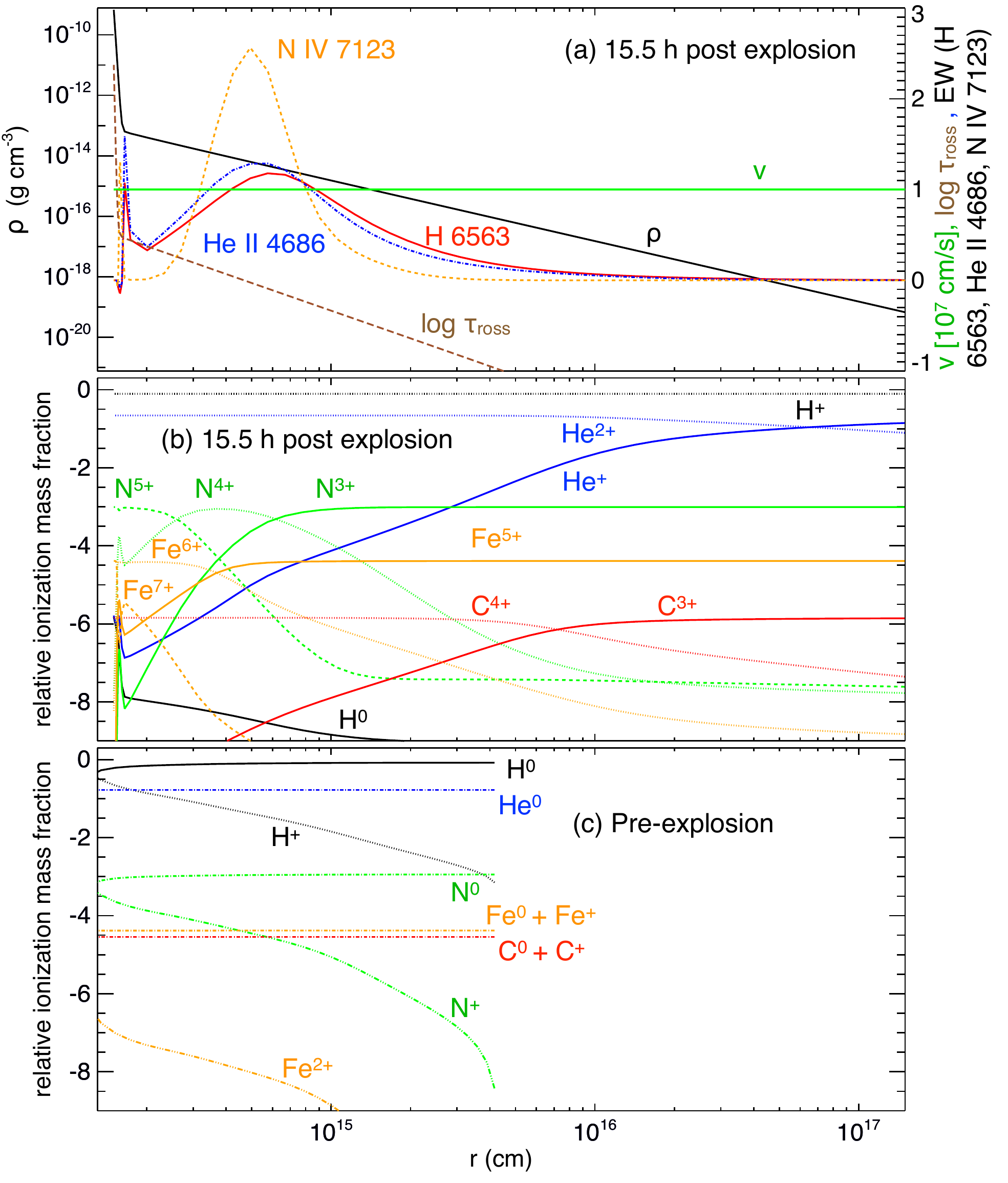}}\\
\caption{\label{fig3}{ (a) Density, velocity, Rosseland optical depth, and formation regions of H$\alpha$, \ion{He}{ii} $\lambda$5411, and \ion{N}{iv}  $\lambda$7123 (area under curve is proportional to the equivalent width) of SN 2013cu based on our CMFGEN model at 15.5h after the explosion. (b) Predicted ionization structure at 15.5 after the explosion. (c) Predicted ionization structure of a possible LBV progenitor before the SN explosion. 
}}
\end{figure}

Figure \ref{fig3}a displays the physical conditions of SN 2013cu at 15.5h after explosion. Notice that the photosphere is indeed located in the progenitor wind. We find that the strongest lines in the 15.5h optical spectrum, such as H$\alpha$, \ion{He}{ii} $\lambda$5411, and \ion{N}{iv}  $\lambda$7123, are formed over an extended distance of $2-20\times10^{14}$~cm, with the bulk of the emission coming from $4-7\times10^{14}$~cm. We also obtain that the outflow is significantly ionized, with H and He being fully ionized up to large distances ($\gtrsim10^{16}$~cm). The ionization structure at 15.5h post-explosion is shown in Fig. \ref{fig3}b.

Although an {\it apparent} WN(h)-like spectrum is seen at 15.5~h after the explosion, we argue here that the spectral type of the progenitor was not the same as that seen during the post explosion. This is because of the very different ionization conditions at the pre- and post-explosion epochs (Fig. \ref{fig3}b,c). Since the number of ionizing photons at the inner boundary at 15.5~h is  many orders of magnitude larger than the typical number of ionizing photons of SN progenitors, the progenitor wind is quickly ionized. Our model shows that the ionization structure after the SN explosion responds to the new SN effective temperature and huge increase in luminosity (Fig. \ref{fig3}b).

Because of the non-stellar origin of the apparent WN(h) post-SN explosion spectrum, we advocate here for the use of the prefix 'X' (short for eXplosion) before the spectral type of such events like SN and SN impostors where a shock breakout and/or interacting CSM illuminate and ionize the precursor wind. This will avoid future confusion between spectral types seen after the explosion with those of their progenitors. Thus, in the case of the SN 2013cu spectrum at 15.5~h, the spectral type would be XWN5(h), following the WN classification scheme of \citet{ssm96} and noting that the \ion{C}{IV} $\lambda$5808/\ion{He}{II} $\lambda$5411 ratio favors the earlier classification.

\section{\label{mdot} An LBV/YHG precursor directly before the SN}

To investigate the spectral morphology of the progenitor, we compute its predicted spectrum using as input the values obtained from the post-explosion spectral modeling ($\mdot \simeq 3\times10^{-3}~\msunyr$, $\vwind \simeq 100~\kms$, and chemical abundances). We use CMFGEN in its standard stellar mode as described in, e.\,g., \citet{ghd09}, where a hydrostatic solution is computed for the subsonic region and merged to the wind solution at 0.75 of the sonic speed. The progenitor wind is assumed to accelerate following a beta-type law with $\beta=1$. Since we have no constraints on the luminosity (\lstar) and hydrostatic radius (\rstar) of the progenitor, we assume typical values of $\rstar=100~\rsun$ and $\lstar=5\times10^{5}$~\lsun.  Owing to the large $\mdot$ and low $\vwind$ inferred from the post-explosion spectrum, our pre-explosion models show that the progenitor has a  pseudo-photosphere with $\teff\simeq8000K$. The qualitative pre-explosion spectral morphology and \teff\ are only weakly dependent on \lstar\ and \rstar\ for sensible changes on these values.  Consequently, we are not able to qualitatively differentiate between progenitor models with $\rstar=10$ or 100~\rsun, i.e. even a compact progenitor would show an LBV morphology immediate before the SN explosion.  A similar situation where the hydrostatic layers are deeply buried in an optically-thick wind is seen in Eta Carinae \citep{hillier01,hillier06,ghm12}. 

Figure \ref{fig1}b shows a prediction of the possible pre-explosion optical spectrum. The pre- and post-explosion spectra differ significantly due to the different $\teff$, luminosity, and ionization structures in the two situations (Fig. \ref{fig3}b,c). This example of pre-explosion spectrum resembles those seen in luminous blue variables (LBV). LBVs are a class of unstable massive stars that are found both as a transitional stage from an O-type to a WR star \citep{hd94,gme14} and as SN progenitors \citep{kv06,smith07,smith11_sn2010jl,kochanek11,pastorello07,trundle08,galyam09,gv11,mauerhan12,ofek13b,gme13,gmg13}. 

Another possibility is that the progenitor was a cool yellow hypergiant (YHG), with an inflated radius of $\rstar\gtrsim10^{13}$~cm and $\teff \simeq 5000-7000~\K$ (see \citealt{gmg13}). This is entirely possible given our lack of constraints on \rstar. The progenitor would be losing mass in an outburst similar to that of the Galactic YHG $\rho$ Cas in the year 2000--2001 \citep{lobel03}, which further strength the link between SN 2013cu and YHGs. Of crucial importance here is that the progenitor of this SN IIb would be a YHG, and not a yellow supergiant (YSG) as often quoted in the literature. This is because YHGs have extensive mass loss and outbursts \citep{dejager98}, which is exactly what we infer for the progenitor of SN 2013cu (see discussion in \citealt{gmg13}).

Presumably, the hydrostatic radius of the progenitor could be constrained by future hydrodynamical modeling of the SN lightcurve, in a similar fashion as it has been done for other SN IIb progenitors \citep{bersten12,bersten14,fremling14}. While this could certainly help to rule out a compact progenitor and further confirm our suggestion, it would be hard to differentiate between LBVs and YHGs, since they both share similar values of radii (up to several $10^{13}$~cm). To separate between LBVs and cool YHGs, constraints on the luminosity and/or effective temperature of the progenitor would be needed, but this is only directly accessible by detecting the progenitor in pre-explosion images. For distant events as in the case of 2013cu ($d=108~$Mpc), however, this would be challenging with the current instrumentation.

Therefore, we argue that the high value of \mdot, relatively low \vwind, and chemical abundance pattern are consistent with the progenitor being either an LBV or YHG (and not a WN star) immediately before the SN explosion. It may be possible that the star had a totally different spectral morphology  {\it before} the LBV/YHG eruption phase. Based on the chemical abundances, at this pre-LBV/YHG eruptive phase, the progenitor could have been a WN, LBV, blue supergiant, YHG, or yellow supergiant, depending on its \rstar\ and \mdot. Time-dependent modeling of the SN 2013cu spectral evolution is warranted to investigate the rapid evolution before core collapse. Concluding, we obtain the following temporal evolution for SN 2013cu and its progenitor:\\

\begin{tabular}{c c c c c}
LBV/YHG & $\rightarrow$ & XWN5(h)  & $\rightarrow$ & SN IIb\\
(pre-SN)&  & (15.5 h post-SN)&  & (69 d post-SN)
\end{tabular}

\section{\label{disc} Prospects with UV spectroscopy and outlook}

\begin{figure}
\resizebox{0.995\hsize}{!}{\includegraphics{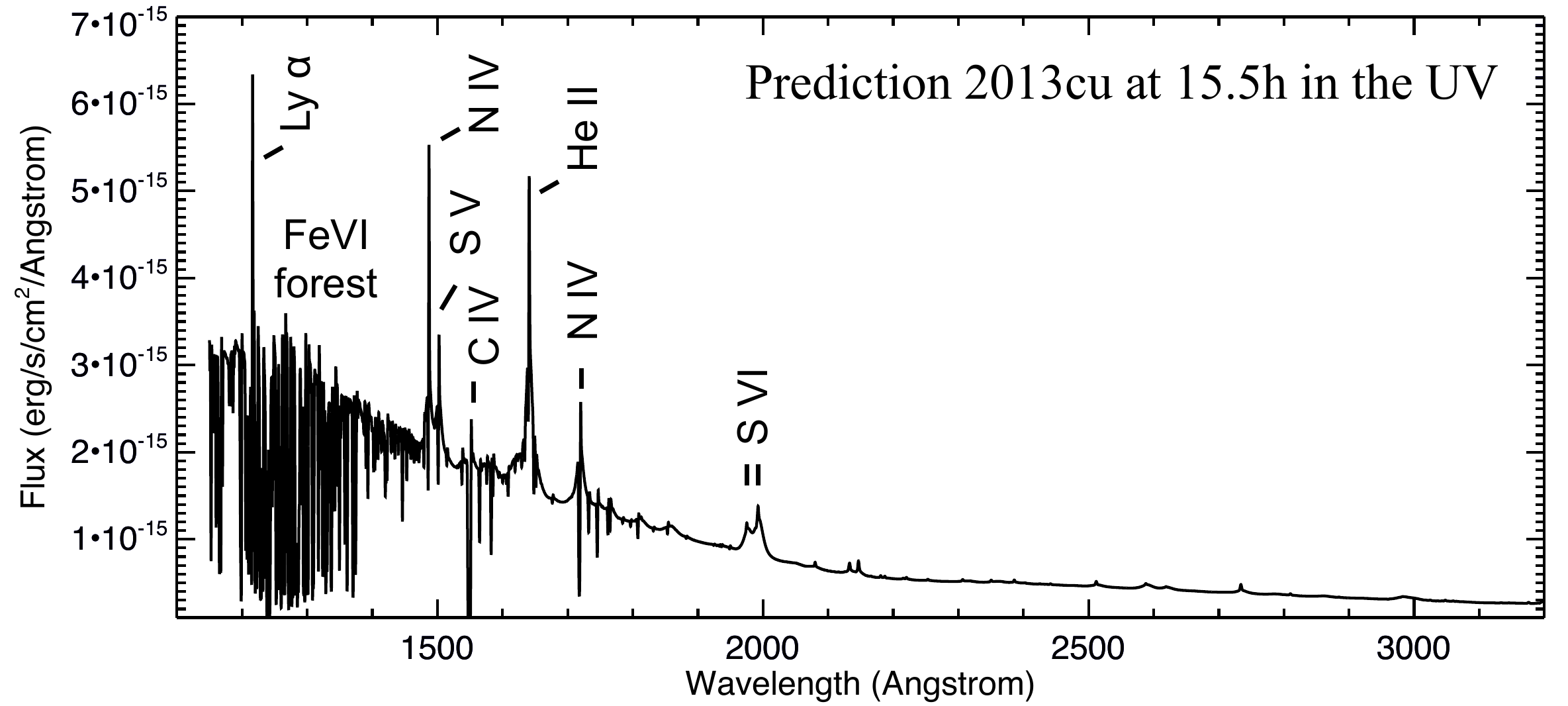}}\\
\caption{\label{fig2}{ Prediction of the early UV spectrum of SN 2013cu based on our CMFGEN model at 15.5h after the explosion. A reddening of $A_\mathrm{V}=0.1$ mag is assumed. }}
\end{figure}


The optical spectra of SN 2013cu, despite having a large wavelength coverage, is relatively poor in bright emission lines. Having access to more spectral lines would significantly aid the determination of the progenitor properties, and this would optimally be achieved by observing in the UV.  

UV spectroscopy offers several other advantages, such as allowing one to observe as close as possible to the peak of the spectral energy distribution of SNe at early epochs. Our predicted spectrum indicates a flux an order of magnitude higher in the UV at 15.5h (Fig. \ref{fig2}). Observing in the UV  allows one to probe resonance lines, such as \ion{C}{iv} $\lambda$1548--1551, and other strong features such as \ion{He}{ii} $\lambda$1640, and  \ion{N}{iv} $\lambda$1718. Given their high optical depths, observing UV resonance lines should provide detectable spectral features even at much lower wind densities than those of SN 2013cu. In addition, these lines allow a more precise determination of the velocity structure of the progenitor wind and SN/wind interacting region, better constraining the models. Our models predict that a forest of \ion{Fe}{vi} and \ion{Fe}{vii} lines should be detectable at $\lambda$$\lambda$ 1200--1450. These have the potential to be used as direct probes of the progenitor and SN metallicities. Notice that, in this particular case,  the predicted spectrum is almost featureless in the range 2000--3000~{\AA}, suggesting that observations should be focused below 2000~{\AA}.

Early spectroscopy of SNe opens up a new observational window that allow the inference of progenitor properties hitherto unavailable, or available only in special cases of strong-interacting SNe. We have showed that when these recent observations are combined  with non-LTE radiative transfer modeling, quantitative information can be obtained about the progenitor such as its chemical composition, wind velocity and mass-loss rate. These quantities allow us to better infer the nature of the progenitor, in particular because a realistic progenitor spectrum can be computed, as illustrated in Fig.~\ref{fig1}b.
 
In a general case, the apparent spectral morphology at early times will change according to the progenitor wind density and composition, time after SN shock breakout,  and SN luminosity. We predict that this will produce a variety of spectral morphologies as the number of events with early-time spectra increases.  

For the new events, early-time SN spectroscopy has the potential to provide even more constraints on the progenitor properties if combined with hydrodynamical modeling of the SN lightcurve and/or fortuitous pre-explosion imaging of the progenitor.  Having progenitors detected in pre-explosion imaging {\it and} spectroscopically observed within a day after the SN explosion will allow us to put definitive constraints on their luminosities, effective temperatures, chemical abundances, wind velocities and mass-loss rates. This will be a key step towards a more complete understanding of the diversity of  SN types and the evolutionary channels that produce their progenitors.

\begin{acknowledgements}
JHG is supported by an Ambizione Fellowship of the Swiss National Science Foundation. JHG thanks A. Gal-Yam, O. Yaron, and E. Ofek for making the observed spectrum of SN 2013cu available, valuable discussions, and the warm hospitality at the Weizmann Institute of Science (Israel). Discussions with J. Hiller and T. Moriya are also acknowledged.
\end{acknowledgements}

\bibliographystyle{aa}
\bibliography{../refs}

\end{document}